# Quantum Gravity and the Origin of the Universe

## -- an Ephemeralist's View

—

B. L. Hu, *Professor of Physics, University of Maryland, U.S.A*

*-- Lecture at the HK University of Science and Technology on May 24, 2006, Hong Kong*

2006  5  24

—

# 1. Origin of the Universe and Quantum Gravity

*1.1 What is meant by the `Origin' of the Universe?*

Based on observational evidence the general belief today is that our universe has been expanding, is adequately described by the Standard Model (based on the Robertson-Walker metric with a Friedmann solution) and had undergone a rapid exponential expansion much earlier, depicted by the inflationary model (Guth 1981). From Einstein's equations this implies that at a finite time in the past the universe must have been in a state of ultra-high curvature and mass density. This is what is generally referred to as the `Big Bang' (BB). The theorists (Penrose, Hawking, Geroch 1967) who found it to be unavoidable in general relativity (GR) theory called this the cosmological `singularity'.

Now, what exactly is meant by the `origin' of the Universe? How did it come into being? Could it be avoided? And, more provocatively, what is *before the Big Bang*? These are the ultimate questions human beings are privileged enough to ask and to think. To address these questions requires some knowledge about the state, structure and dynamics of spacetime itself. We need a theory of the microscopic structures of spacetime.

The laws of classical gravity are believed to be valid from the scale of super-clusters of galaxies to the Planck scale, at $10^{-33}$ cm or $10^{-43}$ sec, an exceedingly small scale indeed. Smaller than this or earlier, we need to apply the laws of quantum physics, thus ushering in quantum gravity theory (QG), which for 30 years before modern string theory was born has been at one of the most challenging frontiers of theoretical physics. For most of that time, quantum gravity research focused on finding ways to quantize GR. The most developed theory from this vein is the loop quantum gravity program. The competing superstring theory has a very different character and most adherents believe that they have already found such a theory. (You can find out more about string theory from Professor Tye's talk.)

*1.2 What is Quantum Gravity? -- Classical to Quantum, Macro to Micro*

The handful of schools may differ in their approaches to quantum gravity (see reading list), but they are likely to agree on one common agenda: The goal of quantum gravity is to uncover the microscopic structure of spacetime. However, the way quantum gravity is defined or how the search is conducted varies a lot: Some believes that quantizing the macroscopic variables of spacetime (metric or connection) will yield the theory of microscopic structures. This has been pursued by general relativists for more than half a century. To others, spacetime is made up of strings or loops. Their tasks are to show how the familiar spacetime structure as we know it today arose. We can refer to this class of theories as the `top down' models. Here I want to mention a third school of thought, which views the macroscopic variables of spacetime as *derived, collective* variables valid only at very low energies and large scales but loses meaning all together at much higher energies or smaller scales (e.g., smaller than the Planck length). This view advocates that one should forget about quantizing these macroscopic variables but search for the microscopic variables directly. Let me describe a view I favor.

*1.3 General Relativity as the Hydrodynamics of Spacetime Microstructure.*

In this school of thought which begun with Sakharov in 1968, general relativity is viewed as the hydrodynamic (the low energy, long wavelength) regime of a more fundamental microscopic theory of spacetime, and the metric and the connection forms are the *collective variables* derived from them. At shorter wavelengths or higher energies, these collective variables will lose their meaning, much as the vibrational modes of a crystal will cease to exist at the atomic scale. If we view GR as hydrodynamics and the metric or connection forms as hydrodynamic variables, quantizing them will only give us a theory for the quantized modes of collective excitations, such as phonons in a crystal, but not a theory of atoms or quantum electrodynamics (QED) which is a more fundamental theory.

According to this view most macroscopic gravitational phenomena can be explained as collective modes and hydrodynamic excitations, from gravitational waves as weak perturbations, to black holes in the strong regime, as solitons. With better observational tools or numerical techniques available, we may even find analogs of turbulence effects in this geometro-hydrodynamics.

This view is very different in meaning and in practice from the other approaches. To better appreciate the roots of these differences, let us step back for a moment and review the two major paradigms in physics which underscore the two main directions of research in theoretical cosmology.

**2. Two major paradigms in Physics and Two directions of Research in Cosmology**

There are two basic aspects in the formulation of any cosmological model. One aspect involves the basic constituents and forces, the other involves the structure and dynamics, i.e., the organization and processing of these constituents as mediated by the basic forces or their derivatives. The first aspect is provided by the basic theories describing spacetime and matter. The second aspect in addressing the universe and its constituents is cosmology proper.

*2.1 Two paradigms in Physics: Elementary Particle versus Condensed Matter Physics*

It is not difficult to recognize that actually these two aspects permeate throughout almost all subfields in physics, or science in general. Examples of the first aspect in physics dealing with the "basic" constituents and forces are general relativity, quantum field theory, quantum electrodynamics (QED), quantum chromodynamics (QCD), grand unified theories, supersymmetry, supergravity and superstring theories. The second aspect dealing with the structure and dynamics is the subject matter of biology, chemistry, molecular physics, atomic physics, nuclear physics and particle physics. The former aspect is treated today primarily in the disciplines of elementary particle physics and quantum gravity. The latter aspect is treated today in the discipline known collectively as condensed matter physics. In this sense we can, for example, regard nuclear physics as the condensed matter physics of quarks and gluons.

Note, however, the *duality* and the *interplay* of these two aspects in any discipline. On the one hand the basic laws of nature are often discovered or induced from close examination of the structure and properties of particular systems - witness the role played by atomic spectroscopy and scattering in the discovery of quantum mechanics and atomic theory, accelerator experiments in advancing particle physics. On the other hand, once the nature of the fundamental forces and constituents are known, one attempts to depict reality by deducing possible structures and dynamics from these basic laws. Thus the study of electrons and atoms via electromagnetic interaction has been the underlying fabric of condensed matter physics for a while. Deducing nuclear force from QCD remains the central task of nuclear research today. From general relativity one attempts to deduce the properties of neutron stars, black holes and the universe, which is the theme of relativistic astrophysics and cosmology.

Note that many known physical forces are not *fundamental* (in the sense that they are irreducible), but are *effective* in nature. Molecular forces and nuclear forces are such examples. One may also regard gravity as an effective force. Note also that many disciplines contain dual aspects. This is especially true in the developing areas, in which the basic forces and constituents of the system are not fully understood. For example, particle physics deals both with the structure and the interactions as quantum flavor and color dynamics. The duality of compositeness and elementarity should also be present in superstring and many quantum gravity theories.

*2.2 Two Aspects in Cosmology*

What about cosmology? The above-mentioned dual aspects are certainly apparent. What is new is that in addition to matter (as described by particles and fields) we have to add in the consideration of spacetime (as described by geometry and topology). In the first aspect concerning constituents and forces, there are also two contrasting views. The "idealist" takes the view that spacetime is the basic entity, the laws of the universe is governed by the dynamics of geometry. Matter is viewed as perturbations of spacetime, particle as geometrodynamic excitons. These ideas, as extension of Einstein's theory, are not that strange as they may appear: Particles are representations of internal symmetries, graviton the resonant modes of strings. By contrast, the "materalist" takes the view that spacetime is the manifestation of collective, large scale interaction of matter fields. Thus according to Sakharov, gravity should be treated as an effective theory, like elasticity to atomic forces. This is expressed in the induced gravity program. Despite its many technical difficulties, this view still evokes some sobering thoughts. It suggests among others that the attempt to deduce a quantum theory of gravity by quantizing the metric may prove to be as meaningful as deducing QED from quantizing elasticity.

In recent years the apparent contrast between particle-fields and geometry-topology has dissolved somewhat in the wake of superstring theory. The fact that the same concept can be viewed in both ways and that spacetime and strings appear in different regimes may indeed offer some new insights into the fundamental aspects of our universe. Duality between the high and low energy sectors, correspondence between gauge theory in the bulk and conformal field theory on the boundary and the holography principle of meaningful information residing on the surface are perhaps some of the most attractive ideas evolved.

As for the second aspect in cosmology, i.e., the manifestation of basic forces in astrophysical and cosmological processes, one sees that almost with any subdiscipline of physics there is a corresponding branch of astrophysics. However, the central theme of cosmology which addresses the state of the universe as a whole is more than the sum-total of its individual components, as depicted by the many subdisciplines of astrophysics. There are broader issues special to the overall problem of how the universe comes into being and why it should be the way it is, which may be traced back to the puzzling issues at the foundation of quantum mechanics and general relativity. The inquiry of these issues will necessarily bring us back to the `fundamental' aspect of inquiries discussed above.

*2.3 Two Directions of Cosmological Research*

Depending on the relative emphasis one puts on these two aspects, current research on cosmological theories follow roughly two directions:

A) Cosmology as consequences of quantum gravity and superstring theories.
B) Cosmology describing the structure and dynamics of the universe

In this first direction one could also include inquiries or proposals made which view the universe as manifestor of physical laws, as formulator of rules, as processor of information, etc. This direction of cosmological research touches on the basic laws of quantum mechanics, general relativity and statistical mechanics. In this field the formulation of meaningful problems are almost as important as seeking their solutions. Progress will be slow but the intellectual reward is profound.

The second direction is characterized in my figurative depiction "Cosmology as `Condensed Matter' Physics", the title of a paper for a conference held in Hong Kong in 1987. By "condensed state" I refer both to matter and spacetime. Cosmology is the study of the organization and processing of matter as well as spacetime points. I presented via several tables in that article some major ingredients of condensed matter physics, nuclear physics and the physics of the early universe and outlined the major themes of recent development of condensed matter physics. Notice the increasing importance attached to nonlinear, nonlocal and stochastic behavior of complex systems. In my opinion, two new ingredients will likely play a dominant role in the structure and evolution of the primordial universe at the Planck scale: One is topology and the other is stochasticity, both for matter-field and spacetime-geometry.

New impetus is provided by advances both in 1) particle physics and quantum gravity, such as superstring theory, loop quantum gravity and simplicial gravity, where close mathematical formulation of these problems has become possible, and 2) condensed matter physics such as critical dynamics and quantum phase transition, order-disorder cross-over, dynamical and complex systems, etc. Developing the synergy between these two major disciplines or sectors of physics can open up new possibilities in probing the organization and dynamics of matter in various states. These techniques and ideas may also provide useful hints in understanding how spacetime takes shape, how the universe evolves, what determines its content and how its many different structural forms develop. Cosmological research would benefit from recognizing and harnessing these resources.

### 3. The Mescoscopic Structures of Spacetime and Stochastic Gravity

As remarked in Sec. 2, our view of quantum gravity is that we find it more useful to find the micro-variables than to quantize macroscopic variables. And of the two paradigms described above, in order to unravel the microscopic structure of spacetime, our outlook is closer to the condensed matter than the elementary particle physics paradigm. Our approach relies more on statistical and stochastic methods and our focus will be on how the known macroscopic levels of structure emerge from the unknown underlying substructures. If we view classical gravity as an effective theory, i.e., the metric or connection functions as collective variables of some fundamental constituents which make up spacetime in the large, and general relativity as the hydrodynamic limit, we can also ask if there is a regime like kinetic theory of molecular dynamics or mesoscopic physics of quantum many body systems intermediate between quantum micro-dynamics and classical macro-dynamics. In addition to serving many practical applications, mesoscopic physics also embodies some fundamental issues. It dwells on two central issues in theoretical physics: the micro to macro and the quantum to classical transitions. To identify any intermediate levels of structure of spacetime between the macro and the micro, it is useful to examine the existing gravitational theories beginning with GR.

**3.1** *The three lowest layers: Classical, Semiclassical and Stochastic Gravity*

The theory of general relativity provides an excellent description of the features of large scale spacetime and its dynamics. *Classical gravity* assumes classical matter as source in the Einstein equation. When quantum fields are included in the matter source, a quantum field theory in curved spacetimes is needed. At the semiclassical level the source in the Einstein equation is given by the expectation value of the energy momentum tensor operator of quantum matter fields with respect to some quantum state. *Semiclassical gravity* refers to the theory where classical spacetime is driven by quantum fields as sources, thus it includes the backreaction of quantum fields on spacetime and the evolution of quantum field and spacetime self-consistently. This is the theory where fundamental discoveries in black hole physics by example of Hawking radiation and in cosmology such as the inflationary universe were made. This serves as a solid platform for us to start the expedition towards quantum gravity, which emphasized, is a theory of the microscopic structure of spacetime, not necessarily obtainable from quantizing general relativity. The next higher layer is *stochastic gravity* which includes the fluctuations of quantum field as source described by the Einstein-Langevin equation. Our approach to *quantum gravity* uses stochastic gravity as a launching platform and mesoscopic physics as a guide.

So what is the physics of mesoscopic spacetime and how does stochastic gravity enter?

**3.2** *Mesoscopic Structure and Stochastic Gravity*

In a 1994 conference paper I pointed out that many issues special to this intermediate stage between the macro and micro structures, such as the transition from quantum to classical spacetime via the decoherence of the `density matrix of the universe', phase transition or cross-over behavior at the Planck scale, tunneling and particle creation, or growth of density contrast from vacuum fluctuations, share some basic concerns of mesoscopic physics in atomic/optical, particle/nuclear and condensed matter or quantum many body systems.  Underlying these issues are three main factors: quantum *coherence, fluctuations and correlations*.  We discuss how a deeper understanding of these aspects of fields and spacetimes related to the quantum / classical and the micro / macro interfaces, the discrete / continuum or the stochastic / deterministic transitions can help to address some basic problems in gravity, cosmology and black hole.

Stochastic gravity is a consistent and natural generalization of semiclassical gravity to include the effects of quantum fluctuations. The centerpiece of this theory is the stress-energy bi-tensor and its expectation value known as the noise kernel. We believe that precious new information resides in the two-point functions and higher order correlation functions of the stress energy tensor which, through the generalized Einstein-Langevin equations for the higher-order induced metric correlations, may reflect the finer structures of spacetime at a scale when information provided by its mean value as source (semiclassical gravity) is no longer adequate. The key point here is the important role played by noise, fluctuations, dissipation, correlations and quantum coherence, the central issues focused on by mesoscopic physics. Noise carries information about the correlations of the quantum field and the quantum coherence in the gravity sector is obtained from the correlations of induced metric fluctuations. Stochastic gravity provides a relation between noise in quantum fields and metric fluctuations.

This new framework allows one to explore the quantum statistical properties of spacetime: How fluctuations in the quantum fields induce metric fluctuations and seed the cosmic structures, quantum phase transition in the early universe, black hole quantum horizon fluctuations, stochastic processes in the black hole environment, the back-reaction of Hawking radiance in black hole dynamics, and implications on trans-Planckian physics. On the theoretical issues, stochastic gravity is the necessary foundation to investigate the validity of semiclassical gravity and the viability of inflationary cosmology based on the appearance and sustenance of a vacuum energy-dominated phase. It is also a useful platform supported by well-established low energy (sub-Planckian) physics to explore the connection with high energy (Planckian) physics in the realm of quantum gravity.

**3.3** *Spacetime as an Emergent Collective State of Strongly Correlated Systems*

Viewing the issues of correlations and quantum coherence in the light of mesoscopic physics we see that what appears as a source in the Einstein-Langevin equation, the stress-energy two point function, is analogous to conductance of electron transport which is given by the current-current two point function. What this means is that we are really calculating the transport functions of the matter particles as depicted here by the quantum fields. Following Einstein's observation that spacetime dynamics is determined by (while also dictates) the matter (energy density), we expect that the transport function represented by the current correlation in the fluctuations of the matter energy density would also have a geometric counterpart and equal significance at a higher energy than the semiclassical gravity scale. This is consistent with general relativity as hydrodynamics: conductivity, viscosity and other transport functions are hydrodynamic quantities. Here we are after the transport functions associated with the dynamics of spacetime structures. The Einstein tensor correlation function in Minkowsky spacetimes calculated by Martin and Verdaguer is a first step. Another example is given by Shiokawa who computed the metric conductance fluctuations.

For many practical purposes we don't need to know the details of the fundamental constituents or their interactions to establish an adequate depiction of the low or medium energy physics, but can model them with semi-phenomenological concepts. When the interaction among the constituents gets stronger, or the probing scale gets shorter, effects associated with the higher correlation functions of the system begin to show up. Studies in strongly correlated systems give revealing examples. Thus, viewed in the light of mesoscopic physics, aided by stochastic gravity, we can begin to probe into the higher correlations of quantum matter and with them the associated excitations of the collective modes in geometro-hydrodynamics, the kinetic theory for the meso-dynamics of spacetime, and eventually quantum gravity -- the theory of micro-spacetime dynamics.

In seeking a clue to the micro theory of spacetime from macroscopic constructs, we have focused here on the kinetic / hydrodynamic theory and noise / fluctuations aspects. Statistical mechanical and stochastic / probability theory ideas will play a central role. We will encounter nonlinear and nonlocal structures (nonlocality in space, nonMarkovian in time) in abundance. Another equally important factor is topology: Topological features can have a better chance to survive the coarse-graining or effective / emergent processes to the macro world and can be a powerful key to unravel the hidden structures of the microscopic world.

**4. One Vein of the `Hydro' View: Spacetime as Condensate?**

Viewing our spacetime as a hydrodynamic entity, I'd like to explore with you a new idea inspired by the development of Bose-Einstein Condensate (BEC) physics in recent years. The idea is that, maybe spacetime, describable by a differentiable manifold structure, valid only at the low-energy long-wavelength limit of some fundamental theory, is a condensate. We have examined what a condensate means in a recent essay, but for now we can use the BEC analog and think of it as a collective quantum state of many atoms with macroscopic quantum coherence.

4.1 <u>Unconventional view 1</u>: *All sub-Planckian physics are low temperature physics*

Atom condensates exist at very low temperatures. It takes novel ways of cooling the atoms, many decades after the theoretical predictions, to see a BEC in the laboratories. It may not be too outlandish to draw the parallel with spacetime as we see it today, because the present universe is rather cold (~3K). But we believe that the physical laws governing today's universe are valid all the way back to the GUT (grand unification theory) and the Planck epochs, when the temperatures were not so low any more. Any normal person would consider the Planck temperature $T^{Pl}= 10^{32}$K a bit high. Since the spacetime structure is supposed to hold (Einstein's theory) for all eras below the Planck temperature, if we consider spacetime as a condensate today, shouldn't it remain a condensate at this ridiculously high temperature?

YES is my answer to this question. What human observers consider as high temperature (such as that when species homo-sapiens will instantly evaporate) has no effect on the temperature scales defined by physical processes which in turn are governed by physical laws. Instead of conceding to a breakdown of the spacetime condensate at these temperatures, one should push this concept to its limit and not be surprised at the conclusion that all known physics today, as long as a smooth manifold structure remains valid for spacetime, the arena where all physical processes take place, are low-temperature physics. Spacetime condensate began to take shape at the Planck temperature, but will cease to exist above it, according to our current understanding of the physical laws. In this sense spacetime physics as we know it is low temperature hydrodynamics, and, in particular, today we are dealing with ultra-low temperature physics, similar to superfluids and BECs.

4.2 <u>Unconventional view 2</u>: *Spacetime is, after all, a quantum entity*

An even more severe difficulty in viewing spacetime as a condensate is to recognize and identify the quantum features in spacetime as it exists today, not at the Planck time. The conventional view holds that spacetime is classical at scales larger than the Planck length, but quantum if smaller. That was the rationale for seeking a quantum version of general relativity, beginning with quantizing the metric function and the connection forms. The spacetime condensate view holds that the universe is fundamentally a quantum phenomenon, but at the mean field level the many body wave functions (of the micro-constituents, or the `atoms' of spacetime) which we use to describe its large scale behavior (order parameter field) obey a classical-like equation, similar to the Gross-Pitaevsky equation in BEC, which has proven to be surprisingly successful in capturing the large scale collective dynamics of BEC, until quantum fluctuations and strong correlation effects enter into the picture.

Could it be that the Einstein equation depicting the collective behavior of the spacetime quantum fluid is on the same footing as the Gross-Pitaevsky equation for BEC? The deeper layer of structure is ostensibly quantum, it is only at the mean field level that the many-body wave function is amenable to a classical description. We have seen many examples in quantum mechanics where this holds. For any quantum system which has bilinear coupling with its environment or is itself Gaussian exact (or if one is satisfied with a Gaussian approximation description) the equations of motion for the expectation values of the quantum observables have the same form as its classical counterpart. Ehrenfest theorem interconnecting the quantum and the classical is an example.

The obvious challenge is, if the universe is intrinsically quantum and coherent, where can one expect to see the quantum coherence phenomena of spacetime? Here again we look to analogs in BEC dynamics for inspiration, and there are a few useful ones, such as particle production in the collapse of a BEC (Bosenova) experiment. One obvious phenomenon staring at our face is the vacuum energy of the spacetime condensate, because if spacetime is a quantum entity, vacuum energy density exists unabated for our present day late universe, whereas its origin is somewhat mysterious for a classical spacetime in the conventional view.

**5. Implications for the Origin of the Universe and other issues.**

So what does this alternative viewpoint of quantum gravity say about the important issues of cosmology? Let's begin with the low energy phenomena accessible by our observations today. There are many discussions these days about Lorentz invariance being broken at ultra-high energies, from superstring and other theories. Lorentz symmetry is a well established symmetry of local (Minkowsky) structure of our spacetime, first found in Maxwell's equations of electromagnetism, later adopted as the symmetry of special relativity as the new laws of mechanics which displaces the Galilean symmetry underlying Newton's theory. Minkowsky provided the geometric description of this new theory of spacetime.

In the `hydro' spacetime viewpoint, Lorentz invariance is an emergent symmetry which applies only to our large scale structure of spacetime at this late stage of its evolution, not unlike the many symmetries of fluids which are not apparent at the molecular dynamics level, which, at a finer scale, its structure and dynamics are governed by a different set of symmetries. Lorentz and other symmetries came into being at sub-Planckian energy, when spacetime took on a form continuous and smooth enough so that a manifold structure appears and differential geometric description became applicable. At sub-Planckian length it could be in a foam-like structure with non-trivial topologies, as captured in the spacetime foam idea of Wheeler. Lorentz as well as many symmetries associated with a smooth large scale manifold structure will be completely out of place. So it is not such a startling thing to imagine giving up many of our much cherished law and order, because our experience is limited to very special conditions. From the emergent viewpoint or from the philosophy of an `ephemeralist' (e.g., Zuang Zhou) no laws are sacrosanct and nothing is for ever.

Another implication of this view is that with a clear vision that there exists fundamental differences in the levels of structure, it can help us identify untenable ideas and not to waste time in meaningless pursuits. Before describing activities involving spacetime and its more elementary constituents, one should explain where/how spacetime structure comes into being. For example, `string cosmology' as a research area appears strange to me: One cannot just write down a metric, throw strings into it and start proposing a cosmological theory that way. How could strings propagate on a structure as yet to be determined by their interactions? One needs to show how strings made up our spacetime, or at the least, identify the particular regime(s) where strings are free to propagate. In fact, despite the often mention of inflation in many papers on string cosmology, the major players of string theory who are honest in reporting their results are likely to agree that they have not been successful in predicting inflation from string theory. Similarly, speculations into pre-Big Bang, if BB signifies the beginning of spacetime, would be useless if it invokes a manifold structure for the background spacetime.

On the constructive aspects, this view could provide a different and perhaps better approach to important issues, such as the mystery of dark energy: Why is the cosmological constant so low (compared to natural particle physics energy scale) today, and so close to the matter energy density (the coincidence issue)? We have also explored the implications of this view on quantum mechanics and general relativity, and its relation to string theory and loop quantum gravity. You can read more about this in my essay on spacetime condensates and Professor Volovik's book.

Finally, what would this view say about the origin of the Universe? I think phase transition is probably a better way to address this question. We live in a low energy (temperature) phase and our description of spacetime is good only at its very long wavelength limit. The phase transition point according to our present understanding of physics could be the Planck energy (remember we said that even slightly below the Planck temperature is considered low temperature). In this view the origin of the universe is really just the beginning of a new low energy phase, not unlike water turning into ice below the freezing point. For creatures incapable of living in a subzero temperature they would regard the freezing point as the beginning of their universe.

Now, what came before that? In this view it is not difficult to imagine there is an epoch before this `beginning', in fact, perhaps many different epochs and many beginnings – no mysticism here. Before the Big Bang, or if you like sensational advertisements, the 'Birth of the (our) Universe', there existed a different phase in the structure of spacetime. We need a different set of variables to describe its basic constituents (not metric or connection), a different language to capture its structure (not differential geometry) and a different set of equations to describe its dynamics (not general relativity). Unlike the summer insects, even if we cannot live that phase we can surely think about its attributes and even devise tools to capture its essence. That is the power of human intellect, will and spirit imbued in the fearless quests of theoretical physics, that which brought you and me together here today.

—

## Readings and References: